\documentclass[prb,twocolumn,superscriptaddress,amsmath,amssymb]{revtex4-2}
\usepackage{graphicx}
\usepackage{babel}
\usepackage{units}
\usepackage{color}
\usepackage[dvipsnames]{xcolor}
\usepackage{eqnarray}
\usepackage{tabularx}
\usepackage{bbm}
\usepackage{float}
\usepackage{soul}
\usepackage{subfigure}
\usepackage{verbatim}
\usepackage{ulem} 

\begin{document}

\title{Magneto-optical response of the Weyl semimetal NbAs:\\ Experimental results and hyperbolic-band computations}
\author{S. Polatkan}
\affiliation{1.~Physikalisches Institut, Universit\"at Stuttgart, 70569 Stuttgart, Germany}
\author{E. Uykur}
\affiliation{1.~Physikalisches Institut, Universit\"at Stuttgart, 70569 Stuttgart, Germany}
\affiliation{Institut für Ionenstrahlphysik und Materialforschung, HZDR, 01328 Dresden, Germany}
\author{J. Wyzula}
\affiliation{LNCMI, CNRS-UGA-UPS-INSA-EMFL, 38042 Grenoble, France}
\author{M. Orlita}
\affiliation{LNCMI, CNRS-UGA-UPS-INSA-EMFL, 38042 Grenoble, France}
\affiliation{Faculty of Mathematics and Physics, Charles University, 121 16 Prague, Czech Republic}
\author{C. Shekhar}
\affiliation{Max-Planck-Institut f\"{u}r Chemische Physik fester Stoffe, 01187 Dresden, Germany}
\author{C. Felser}
\affiliation{Max-Planck-Institut f\"{u}r Chemische Physik fester Stoffe, 01187 Dresden, Germany}
\author{M. Dressel}
\affiliation{1.~Physikalisches Institut, Universit\"at Stuttgart, 70569 Stuttgart, Germany}
\author{A. V. Pronin}\email{artem.pronin@pi1.physik.uni-stuttgart.de}
\affiliation{1.~Physikalisches Institut, Universit\"at Stuttgart, 70569 Stuttgart, Germany}

\date{December 5, 2023}
	
\begin{abstract}

The magneto-optical properties of (001)-oriented NbAs single crystals have been studied in the spectral range from 5 to 150 meV and in magnetic fields of up to 13 T. A rich spectrum of inter-Landau-level transitions is revealed by these measurements. The transitions follow a square-root-like dependence with magnetic field, but the simple linear-band approximation is unable to accurately reproduce the observed behavior of the transitions in applied fields. We argue that the detected magneto-optical spectra should be related to crossing hyperbolic bands, which form the W1 cones. We propose a model Hamiltonian, which describes coupled hyperbolic bands and reproduces the shape of the relevant bands in NbAs. The magneto-optical spectra computed from this Hamiltonian nicely reproduce our observations. We conclude that the hyperbolic-band approach is a minimal model to adequately describe the magneto-optical response of NbAs and that the chiral (conical) bands do not explicitly manifest themselves in the spectra.
\end{abstract}
\maketitle

Weyl semimetals (WSMs)~\cite{Murakami2007, Wan2011, Burkov2011PRL, YanFelser2017, Armitage2018} are probably the most interesting and currently most studied family in the diverse field of topological bulk electronic phases~\cite{Burkov2011PRB, Manes2012, Soluyanov2015, Bzdusek2016, Zhu2016, Bradlyn2016, Yan2017}. WSMs possess chiral electron bands, in which the quasiparticles behave similar to Weyl fermions~\cite{Weyl1929}. The rich spectrum of different electrodynamic properties of WSMs have been intensively studied in the last years~\cite{Lv2021, Pronin2021, Guo2023}. The major attractions for such studies include the presence of the chiral quasiparticles as such and the possibility for a solid-state realization of the chiral anomaly~\cite{Nielsen1981, Son2013}. The unconventional consequences of these phenomena on the material's electrodynamics stimulated intensive theoretical and experimental studies of optical properties of WSMs~\cite{Hosur2012, Ashby2014, Ma2017, Levy2020, Hutt2019}.

\begin{figure}[]
	\includegraphics[width=\linewidth]{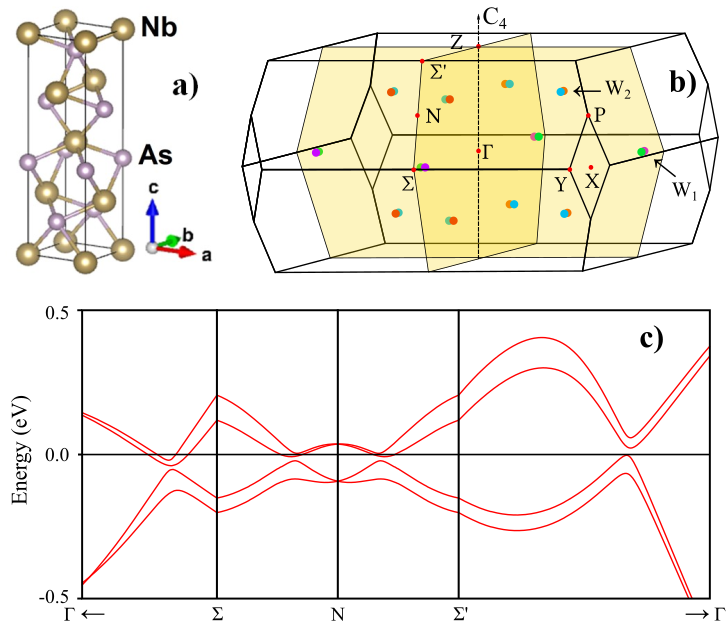}
	\caption{(Color online)
(a) Crystallographic structure, (b) first Brillouin zone, and (c) low-energy DFT band structure of NbAs. The Weyl points of different types and chiralities are shown in different colors and the mirror planes in pale yellow in (b). The small gaps present in (c) are due to the vicinity of the chosen $k$ path to the gapped nodal lines. The W1 points are not far from the $\Gamma$--$\Sigma$ line, while the W2 nodes are in the vicinity of the $\Gamma$--$\Sigma'$ line in (c).}
	\label{NbAs_intro}
\end{figure}

At present, the family of transition metal monopnictides (TaAs, TaP, NbAs, and NbP)~\cite{Weng2015, Lv2015, Xu2015NbAs} is likely the best-known and experimentally most explored class of WSMs. In these nonmagnetic materials, which lack space inversion (space group No. 109, Fig.~\ref{NbAs_intro}(a)) a type-I WSM state is realized. According to band-structure calculations~\cite{Weng2015, HuangSM2015, Lee2015, Grassano2018b}, the Brillouin zone (BZ) of these compounds possesses 24 Weyl nodes, which can be divided in two groups, commonly dubbed as W1 (8 nodes) and W2 (16 nodes), see Fig.~\ref{NbAs_intro}(b). \textit{Ab initio} calculations predict that the Weyl nodes are situated either slightly below or slightly above the Fermi level, giving rise to conical Weyl bands. In addition, trivial parabolic bands are also present. This rather complex low-energy band structure often obscures observation of the effects, related to the chiral carriers.

Magneto-optical Landau-level (LL) spectroscopy in the far-infrared region is a well-established tool for experimental studies of topological materials~\cite{Orlita2014, Akrap2016, ChenZ2017, Shao2019, Polatkan2020, Mohelsky2023}. This method is able to directly probe the inter-LL transitions and to discriminate between parabolic and linear (conical) bands. Hence, complementary to angle-resolved photoemission spectroscopy (ARPES), it probes the material's band structure at low energies, where photoemission often lacks accuracy. Furthermore, the inter-LL spectroscopy may also differentiate between chiral and non-chiral bands, bringing a new piece of experimental information, unavailable to standard ARPES. In the last years, a number of magneto-optical studies on the WSMs form the TaAs family have been published~\cite{Xu2017, Jiang2018, Yuan2018, Yuan2020, Polatkan2020, Wyzula2022, Santos2022, Zhao2022}. These studies generally confirm the presence of the Weyl bands, even though the magneto-optical spectra may look different for different members of the family (also the experiment geometry affects the spectra).

In this paper, we concentrate on one compound -- NbAs -- and study its Landau-level spectrum in the applied external magnetic field in the Faraday geometry. We motivate our study by attempting to consistently describe the major features of the magneto-optical response within a single effective Hamiltonian, which grasps the essence of the band structure. In a previous study~\cite{Yuan2018}, the observed inter-LL transitions in NbAs have been assigned to a number of different simple models, combining linear and parabolic low-energy approximations of the bands for different sections in the BZ. Below, we argue for the validity of our approach.

Single crystals of NbAs were obtained according to the procedure reported in Ref.~\cite{Shekhar2016}: a polycrystalline NbAs powder was synthesized in a direct reaction of pure Nb and As, while the single NbAs crystals were grown from the powder via chemical vapor transport with iodine.

Our magneto-optical data were collected in reflectivity mode in the Faraday configuration from a (001)-oriented facet. The facet area was roughly 2 by 2 mm$^{2}$, and the sample was kept at $T=1.8$~K in helium exchange gas during the measurements. The sample holder was placed in a superconducting coil, which provided magnetic fields up to 13 T. Far-infrared radiation ($\sim5 - 150$ meV) from a globar or Hg lamp was delivered to the sample via light-pipe optics. The reflected light was directed to a Fourier-transform spectrometer and detected by a liquid-helium-cooled bolometer placed outside the magnet. The sample reflectivity $R_{\rm{B}}$ at a given magnetic field $B$ was normalized by the sample’s reflectivity $R_{0}$ measured at $B = 0$.

The \textit{ab initio} calculations were performed with the WIEN2k code~\cite{Blaha2020}, which is based on the (linearized) augmented plane-wave and local orbitals [(L)APW+lo] method to solve the Kohn-Sham equations~\cite{Kohn1965} of density functional theory (DFT). The exchange-correlation potential was calculated using the Perdew-Burke-Ernzerhof generalized gradient approximation~\cite{Perdew1996}. The self-consistent field calculations converged on a $10\times 10\times 10$ $k$-mesh (charge convergence below $10^{-5}~e$, energy below $10^{-6}$ Ry) for the two-dimensional and on a $30\times 30\times 30$ $k$-mesh (charge convergence below $10^{-6}~e$, energy below $10^{-7}$ Ry) for the one-dimensional band structure cuts.

\begin{figure*}[th!]
	\includegraphics[width=\linewidth]{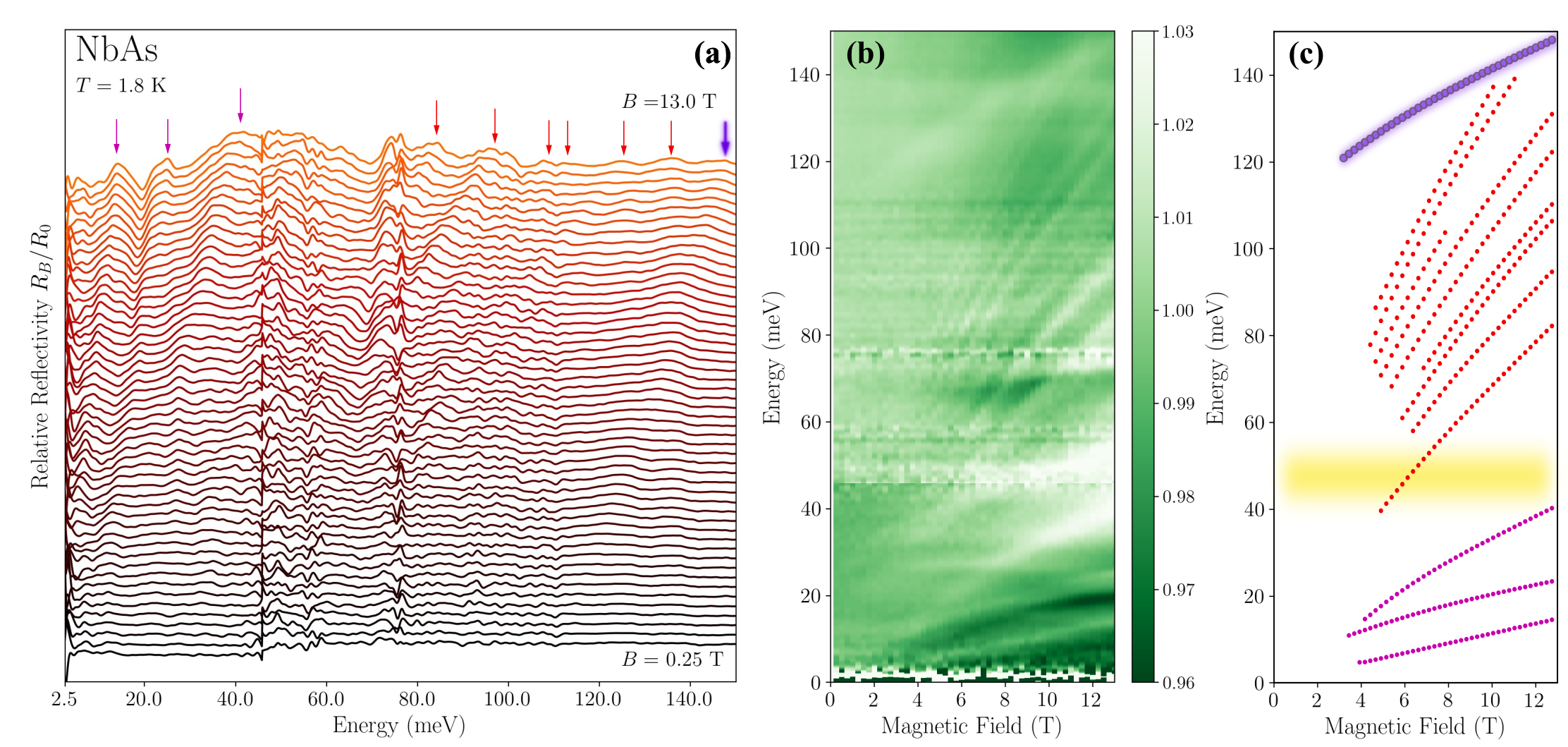}
	\caption{(Color online)
Experimental magneto-optical spectra of (001)-oriented NbAs at 1.8 K. (a) Relative magneto-optical reflectivity, $R_{\rm{B}}/R_{0}$, for $B$ from 0.25 to 13 T in steps of 0.25 T. The spectra are shifted upwards with increasing $B$. The arrows indicate approximate positions of the major transition lines at the highest fields. (b) Same spectra plotted as a false-color plot on the energy--field plane. (c) Schematic diagram showing the observed transitions; the dot colors correspond to the arrow colors in panel (a). The noise level around 50 meV is rather high and no lines can be reliably traced here. This area is shaded yellow, cf. panel (a). (High noise at around 75 meV does not prevent tracing the transition lines.)}
	\label{NbAs_exper}
\end{figure*}

Our main experimental result –- the measured relative reflectivity of NbAs, $R_{\rm{B}}/R_{0}$, is presented in Fig.~\ref{NbAs_exper}~(a,b). A rich series of field-dependent optical features is clearly observable, indicative of inter-LL transitions at $\hbar\omega_{nm}(B)=E_{n}(B) - E_{m}(B)$, where $E_{n}(B)$ are energies of the LLs. We note that the transitions we observe are similar, but not identical, to those reported in Ref.~\cite{Yuan2018}. The exact energy of some of the transitions can differ between our and the previous observations, see Fig.~S1 in the Supplemental Material~\cite{SM}. The detected none-linear variation of $\omega_{nm}$ with $B$, see Fig.~\ref{NbAs_exper}~(b,c), signals a none-parabolic dispersion of the corresponding electronic bands, however a simple conical-band dispersion with its square-root dependence of $\omega_{nm}(B)$ cannot adequately fit the measured spectra either.

Considering the set of transitions in the range 60 to 100 meV marked in Fig.~\ref{NbAs_exper}~(c) by the red dotted lines, one notes that all of these transitions are of similar slope, strength, and width. Hence,
this fan of lines can be associated to inter-LL transitions of a common origin in the band structure. (The other experimental lines will be discussed below.) The fact that these lines are curved, points towards assigning
them to the bands with non-parabolic dispersion, but not necessarily to the transitions between linear bands.

Importantly, the intensity of the inter-LL transitions in 3D bands heavily depends on the out-of-plane velocity $v_z$. This parameter differs by an order of magnitude for the bands near the W1 and W2 nodes: $v_z(\textrm{W}1) \sim 10^4$ m/s and $v_z(\textrm{W}2) \sim 10^5$ m/s \cite{Lee2015, Grassano2018b}. This variation obviously results in a difference in the density of states (DOS) and, via the Kubo formula, in very different intensities of the inter-LL-transition lines, as demonstrated in Fig. S2 of the Supplemental Material~\cite{SM}. In short, the transitions near the W1 cones are expected to dominate over the contributions of the W2 bands in the response of the (001) facet.

We can hence confidently assign the detected magneto-optical features to the band structure that is in proximity to the W1 cones, where $v_z$ is small and the DOS is large. As noted above, it is however impossible to fit the transitions using a single Weyl Hamiltonian with linear bands, also because the observed tight spacing between the inter-LL transitions cannot be reproduced reasonably.

To resolve this issue, we have to construct a model Hamiltonian, which is able to sensibly mimic the DFT band structure and to describe our magneto-optical observations. We now explain the logic we followed in constructing such a Hamiltonian. First, we point out that the intersecting bands in NbAs have a hyperbolic character, cf. Fig.~\ref{NbAs_intro}~(c). Second, these intersecting hyperbolic bands give rise to Dirac nodal lines, if the spin-orbit coupling (SOC) is ignored, see Refs.~\cite{Weng2015, Lee2015}. Thus, we can start building our model with a hyperbolic-band Dirac Hamiltonian. Note, that in our experiments we probe the ab-plane response (which we identify
as the $k_x$-$k_y$ plane in the BZ). Hence, building a 2D model is a sufficient and common approach~\cite{Jeon2014, Jiang2018, Polatkan2020, Zhang2021, Wyzula2022} (the $k_{z}$ dispersion will be included via the DOS,
as mentioned above and discussed in the Supplemental Material~\cite{SM}). Such a Dirac Hamiltonian can be written as:
\begin{equation}
	H_4=
	\begin{pmatrix}
		m \sigma_0 + c_1 \sigma_z + c_2 \sigma_x& \hbar v_x q_x \sigma_x + \hbar v_y q_y \sigma_y \\
		\hbar v_x q_x \sigma_x + \hbar v_y q_y \sigma_y  & -m \sigma_0 - c_1 \sigma_z - c_2 \sigma_x,
	\end{pmatrix}
	\label{H4}
\end{equation}
where the index 4 indicates the matrix size, $m$ is a mass term, $v_{i}$ are velocity parameters, $q_{i}$ are the crystal momenta, $\{\sigma_{i}\}$ are the Pauli matrices, $\sigma_0$ is the identity matrix, and $\{c_{i}\}$ are coupling constants, introduced to provide the band crossings in accordance with the DFT band structure. Without the coupling constants $\{c_{i}\}$, the Hamiltonian describes a gapped Dirac Hamiltonian with its characteristic hyperbolic band dispersion. Similar models have been applied recently to different nodal semimetals~\cite{Koshino2016, Jiang2018, Zhao2022}. As usual, $q_{i} = k_{i}/a_{i}$ with $\{a_{i}\} \equiv \{a, b, c\}$ the lattice constants, $i=\{x,y,z\}$ (recall, NbAs has a tetragonal symmetry), and $k_{i} \in [-\pi, \pi)$. We found that the LLs obtained from the quantized version of Eq.~\eqref{H4} allow one to reproduce the fan of the inter-LL transitions marked with the red dotted lines in Fig.~\ref{NbAs_exper}~(c) quite well,  $m$, $c_{1}$, and $c_{2}$ being around $100 - 150$ meV.

The band crossings obtained in this model, however, are formed by non-degenerate bands. In order to restore the spin degeneracy, we need to double the number of bands and thus we arrive at the following Hamiltonian:
\begin{equation}
	H_8^{\rm{no SOC}}= \begin{pmatrix}
		H_4 & 0 \\
		0 & H_4
	\end{pmatrix}.
\end{equation}
Finally, to model the Weyl nodes, as required by the complete DFT calculations with the SOC included, the $H_4$ Hamiltonians are coupled in the following way:
\begin{equation}
	H_8= \begin{pmatrix}
		H_4 & C \\
		C^\dagger & H_4
	\end{pmatrix}, ~~~ C= \begin{pmatrix}
	0 & c_4 \sigma_x \\
	c_3 \sigma_x& 0
\end{pmatrix}.
	\label{H8}
\end{equation}

\begin{figure}[t]
	\includegraphics[width=\linewidth]{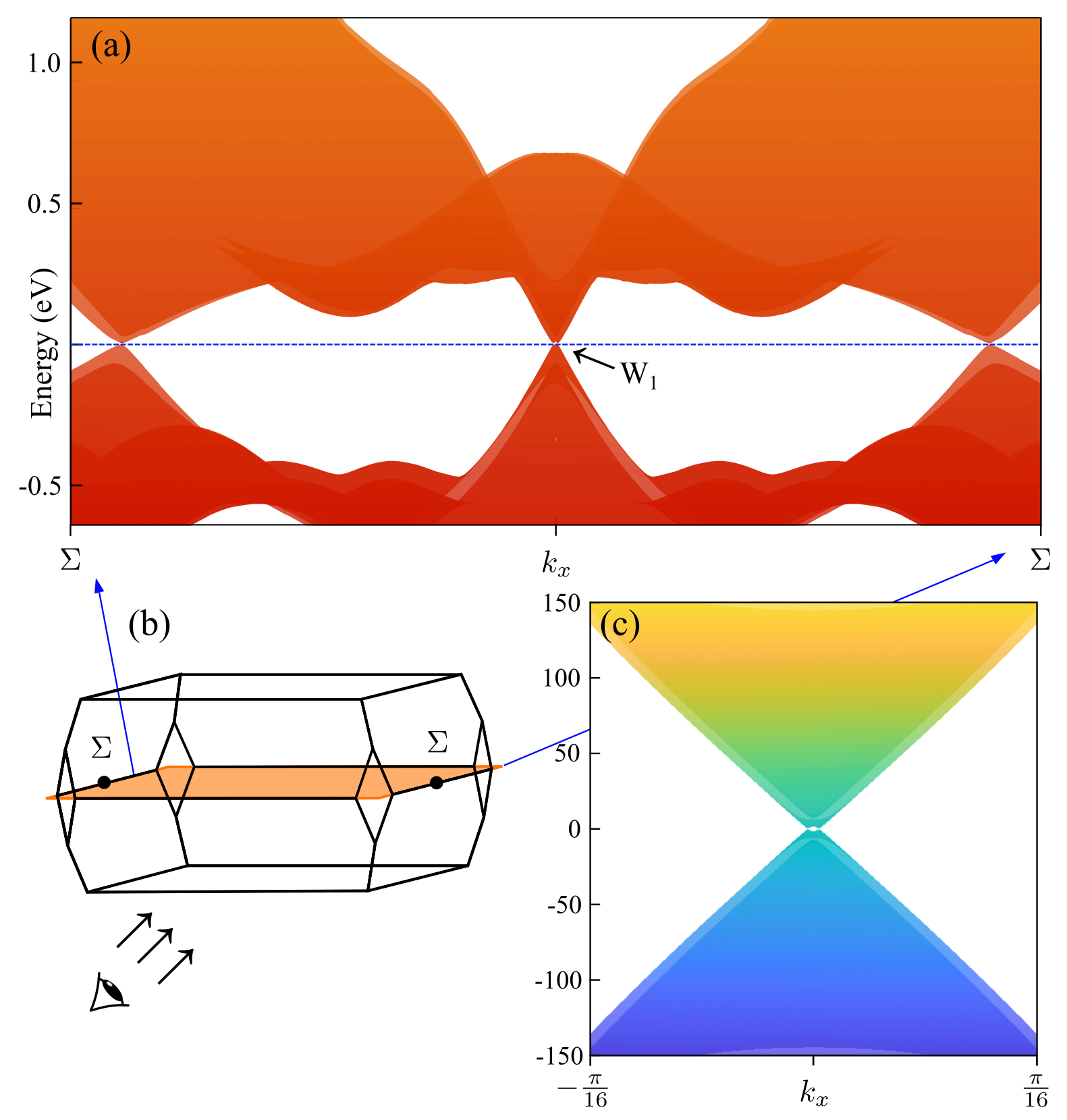}
	\caption{(a) Two-dimensional projection of the NbAs DFT band structure with SOC at $k_z=0$. (b) BZ with the position of the projection plane shown in orange. (c) The bands of the Hamiltonian in Eq.~\eqref{H8} with $c_3=10$~meV and $c_4=-2$~meV, which split the Dirac cones into two Weyl cones along the $k_x$ direction.}
	\label{FigSOC}
	
\end{figure}

In Fig.~\ref{FigSOC}~(a), we provide relevant DFT band structure renderings with the SOC included. It splits the bands such that they overlap by about 15~meV, giving rise to the W1 Weyl cones. Given the strong curvature of the electron band near the point of the intersection, a linear band approximation is questionable within even in a single-digit-meV energy range and also can be obscured entirely by scattering. This indicates that a direct detection of inter-LL transitions within the chiral W1 bands is unlikely in our and similar experiments.

Including the coupling constants in Eq.~\eqref{H8} lifts the band degeneracy and splits the low-energy lines. This allows an accurate description of the lines below 40 meV (magenta dotted lines in Fig.~\ref{NbAs_exper}) within the same model. This way the SOC is effectively included and the pairs of Weyl cones are created from each Dirac cone, see Fig.~\ref{FigSOC}~(c). We note here that the parameters $m$, $c_{1}$, and $c_{2}$ that determine the shape of the spin-degenerate bands (100 meV range, see above) are much larger than the parameters $c_3$ and $c_4$ ($1 - 10$ meV range), which emulate the SOC.

In Fig.~\ref{NbAsLLCalc}, we plot the calculated $R_{\rm{B}}/R_{0}$, which contains the inter-LL transitions obtained from the quantized version of Eq.~\eqref{H8}, which is provided in the Supplemental Material~\cite{SM}. Alongside, the experimentally observed transitions from Fig.~\ref{NbAs_exper}~(c) are presented as dotted lines. The interband optical conductivity was calculated using the Kubo formula. From the conductivity, the relative reflectivity was calculated utilizing the standard optical formulas, see the Supplemental Material~\cite{SM} for details. We found that the best match between experiment and computations can be obtained for: {$v_{x}=5.1 \times 10^{5}~\text{m/s}$, $v_{y}=2.8 \times 10^{5}~\text{m/s}$, $m=179~\text{meV}$, $c_{1}=-132~\text{meV}$, $c_{2}=130~\text{meV}$, $c_{3}=10~\text{meV}$, $c_{4}=-2~\text{meV}$. As one can see from Fig.~\ref{NbAsLLCalc}, the model description of the observed transition lines is very good. Uncertainty estimations of the fit parameters are provided in the Supplemental Material~\cite{SM}. Let us recall that originally we aimed to describe only the transitions marked with the red lines, as they obviously belong to a single band-structure feature. Additionally, if the SOC is taken into account, our model provides a good description of the low-frequency magenta lines as well. Only the high-frequency blurred violet (dotted) line does not fit within the model, but it is obvious that this line cannot belong to the same LL fan, as all other lines. The additional line appearing in the calculations in-between the two experimentally detected sets (i.e., the red and magenta lines) is likely invisible in the experiment because of the large noise at the relevant frequencies, as marked by the yellow area in Fig.~\ref{NbAs_exper}~(c). Although the number of the free parameters in our Hamiltonian may look large, one should recall that these parameters reasonably fit a dozen of observed lines with various field-dependent curvatures and gaps between them. Note that the obtained values of $v_{x}$ and $v_{y}$ are the asymptotic values of the hyperbolic bands and thus represent the upper bonds for the velocity parameters, cf. the values for the Fermi velocities from Refs.~\cite{Lee2015, Grassano2018b}.

\begin{figure}[b]
	\includegraphics[width=0.7\columnwidth]{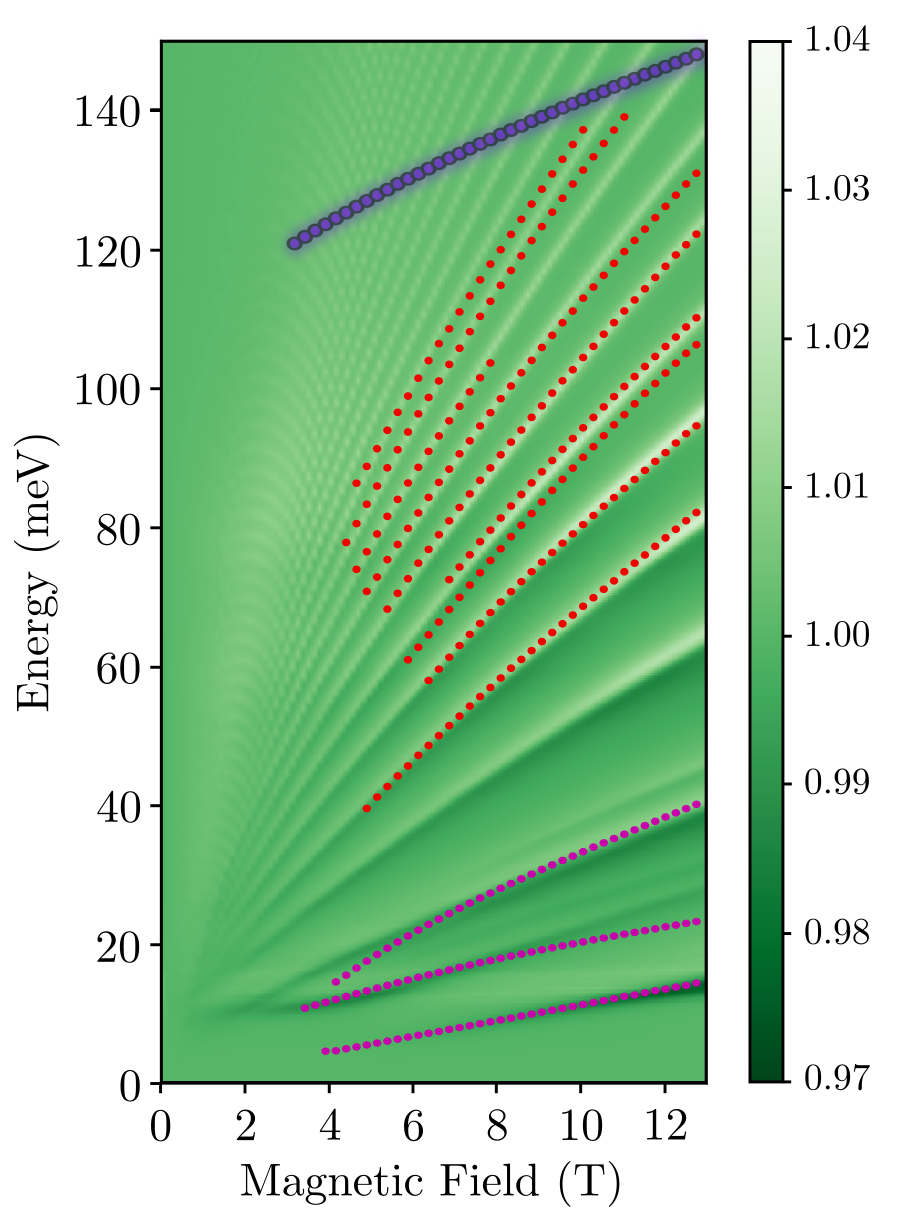}
	\caption{$R_{\rm{B}}/R_{0}$ calculated from the model Hamiltonian Eq.~\eqref{H8} and shown as a false-color plot, in comparison with the experimentally observed transitions, extracted from the magneto-optical data as demonstrated in Fig.~\ref{NbAs_exper}.}
	\label{NbAsLLCalc}
\end{figure}

We conclude the Hamiltonian of Eqs.~\eqref{H8} enables an accurate description of almost all inter-LL transitions observed in our experiment. We also conclude that these transitions belong to a single LL fan and are related to a single band-structure feature -- the crossing hyperbolic bands forming the W1 nodes, as discussed above. One remaining transition (at $120 - 150$ meV, the blurred violet line in Fig.~\ref{NbAs_exper}~(c)) is likely related to gapped bands, in line with the previous assessment~\cite{Yuan2018}.

Summarizing, our results on Landau-level optical spectroscopy demonstrate that the magneto-optical spectra of the Weyl semimetal NbAs at low energies ($5 - 150$ meV) cannot be reduced to such simple approximations, as a linear or parabolic-band response (or a sum of such responses). Instead, the spectra are dominated by the transitions within the crossing hyperbolic bands, which form the W1 Weyl cones. The chiral Weyl bands themselves are not visible, as they have a very small energy scale. Still, taking into account the accurate band structure near the W1 nodes is essential for the proper description of the observed magneto-optical features. The chiral W2 bands, in turn, do not provide detectable contributions due to the low density of states in these bands. Our results are important for correct estimates of the energy scales of chiral carriers in NbAs and for proper interpretations of related results.

\textit{Acknowledgements.} This work was partly funded by the Deutsche Forschungsgemeinschaft (DFG) via Grant No. DR228/51-3 and by a joint German-French Program for Project-Related Personal Exchange by DAAD and ANR via Project No. 57512163.
	
\bibliography{references}

\newpage

\newpage
\vspace*{-2.0cm}
\hspace*{-2.5cm} {
  \centering
  \includegraphics[width=1.2\textwidth, page=1]{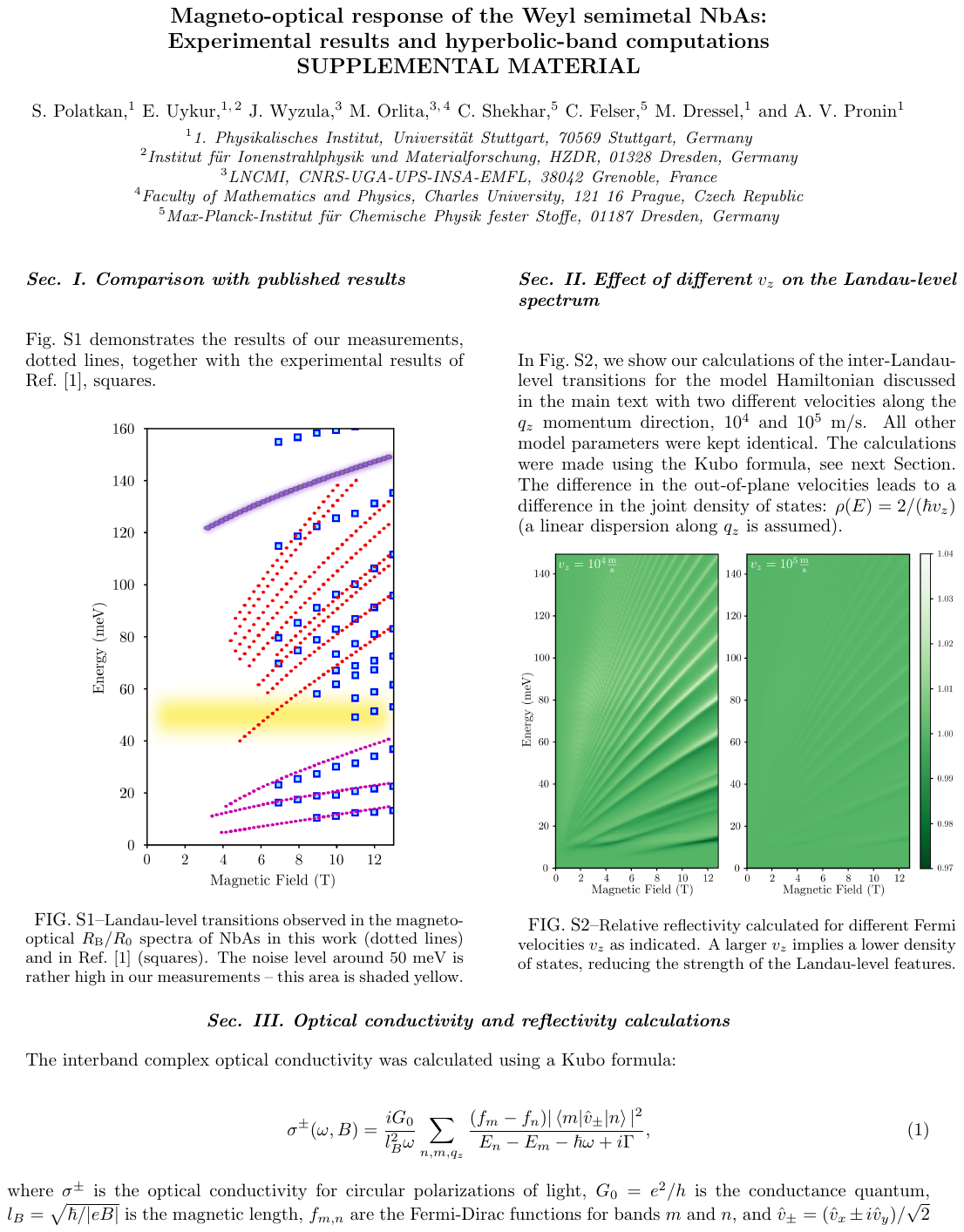} \\
  \ \\
}

\newpage
\vspace*{-2.0cm}
\hspace*{-2.5cm} {
  \centering
  \includegraphics[width=1.2\textwidth, page=2]{sm.pdf} \\
  \ \\
}

\newpage
\vspace*{-2.0cm}
\hspace*{-2.5cm} {
  \centering
  \includegraphics[width=1.2\textwidth, page=3]{sm.pdf} \\
  \ \\
}

\newpage
\vspace*{-2.0cm}
\hspace*{-2.5cm} {
  \centering
  \includegraphics[width=1.2\textwidth, page=4]{sm.pdf} \\
  \ \\
}

\end{document}